\documentclass[preprint,12pt]{elsarticle}
 
\usepackage{amssymb}
\usepackage{amsmath}
\usepackage{color}

\newcommand\Fr{\mbox{\textit{Fr}}}  
\newcommand{\vs}{{\it vs\ }}

\journal{Advances in Water Resources}

\begin{document}

\begin{frontmatter}




\title{Momentum balance in the shallow water equations on bottom discontinuities}

\author[dip]{A. Valiani\corref{Ale}}
\cortext[Ale]{Corresponding author.}
\ead{alessandro.valiani@unife.it}
\author[dip]{V. Caleffi}
\ead{valerio.caleffi@unife.it}
\address[dip]{Universit\`{a} degli Studi di Ferrara,
Dipartimento di Ingegneria, Via G. Saragat, 1 44122 Ferrara, Italy}

\begin{abstract}
This work investigates the topical problem of balancing the shallow water equations
over bottom steps of different heights. The current
approaches in the literature are essentially based on mathematical analysis of the hyperbolic system of balance equations and take into account the relevant progresses in treating the non-conservative form of the governing system in the framework of path-conservative schemes. An important problem under debate is the correct position of the momentum balance closure when the bottom elevation is discontinuous. Cases of technical interest are systematically analysed, consisting of backward-facing steps and forward-facing steps, tackled supercritical and subcritical flows; critical (sonic) conditions are also analysed and discussed.

The fundamental concept governing the problem and supported by the
present computations is that the energy-conserving approach is the
only approach that is consistent with the classical shallow water
equations formulated with geometrical source terms and that the momentum
balance is properly closed if a proper choice of a conventional depth
{\it on\ } the bottom step is performed. The  depth on the step is
shown to be included between the depths just upstream and just
downstream of the step. It is also shown that current choices (as
given in the literature) of the depth on (or in front of) the step can
lead to unphysical configurations, similar to some energy-increasing solutions.
\end{abstract}


\end{frontmatter}


\section{Introduction}
\label{sec:Intro} 
The numerical integration of shallow water equations (SWE) with
source terms has been intensively investigated in recent years due to
the significant advances in computational fluid dynamics in finite difference methods, finite volume methods, discontinuous Galerkin methods, and so forth \cite{Caleffi2006, Caleffi2009, Caleffi12, Caleffi2015}. Particular interest has been devoted to discontinuous solutions, typically due to the physics of stationary jumps and moving bores generated by the hyperbolic nature of the homogeneous problem, in light of the more general Rankine-Hugoniot theory of inviscid shock propagation. A further source of discontinuity is due to the possible discontinuous profile of the bottom elevation, which introduces singularities in the term related to the topography.

From recent works on proper numerical methods for SWE on discontinuous
bed elevations, the first contribution may be considered
\cite{alcrudo01}. The authors analysed the similarity solutions of the
Riemann problem over a step, evidencing the existence of a standing
discontinuity over the step and studying a relevant number of possible
solutions. They imposed total head conservation across the step and
highlighted the need for an additional kinetic energy term to properly
take the energy dissipation into account, when necessary. This work
has the merit of highlighting the wide variety of solutions that can
be found, which are considerably more numerous than the classical case
of the Riemann problem over a flat horizontal bottom. Moreover, this work demonstrates the potential complexity of any numerical method that has the ambition to incorporate the variability of such a number of situations.

The study \cite{LeFloch2011}, which is a deepening of the previous study \cite{LeFloch2007}, investigated the Riemann problem over a
discontinuous bed elevation and provided a detailed analysis of the
possible solutions. This work is likely the most complete study on
this topic. More specifically, the authors investigated the existence
and uniqueness conditions for such solutions, showing the uniqueness
in the non-resonant regime and the existence of multiple solutions in
the resonant regime. The eigensystem analysis provides two genuinely
hyperbolic characteristic fields and one linearly degenerate
field. Regarding the linearly degenerate field, it is shown that
across a discontinuity, the bottom elevation must remain constant or
the discontinuity must be stationary. This result means that over a discontinuous bed, the propagation celerity must be zero. The jump relations on the discontinuity provide the constancy of the unit discharge and the conservation of the total head of the flow, which are exactly the assumptions adopted herein.

The investigation is extended in \cite{HW2014}, where the exact Riemann solutions for the shallow water equations with a bottom step is completed, including the dry bed problem, which can occur in specific circumstances.

The conservation of the discharge and of the total head are successfully applied in 1D modelling of open-channel hydraulics, both for the unit width channel and in a channel of arbitrary shape (\cite{Murillo2013, Murillo2014, Navas2015}).

A different approach is presented in \cite{Bernetti2008}, which presents
the exact solution of the Riemann problem for shallow water equations
with a step-like bottom. The solution was obtained by solving the
system that included an additional equation for the bottom geometry
and then using the principles of conservation of mass and momentum
across the step. By writing the Rankine-Hugoniot condition for the
standing wave across the step, it is shown that the resulting solution
is unique and satisfies the principle of the dissipation of energy
across the shock wave. Because all the methods establish the momentum
integral principle across the step, the key step is estimating
the force exerted by the step on the fluid. In the framework of SWE
and therefore assuming a hydrostatic distribution of pressure, it is
possible to express the step force as a function of an equivalent,
conventional depth {\it on\ } the step. In the case of a forward-facing
step, such a depth is evaluated, as in all the other similar methods,
using the depth just in front of the step measured from the centroid
of the step. In the case of a backward-facing step, which is not explicitly
described, it can be argued that the depth must be evaluated again as
the depth just in front of the step measured from the centroid of the
step. More specifically, the phrase "in front of" means the upstream
flow depth minus half the step height in the forward-facing step case,
and it means the downstream flow depth minus half the step height
depth in the backward-facing step case. This choice is similar to the other methods described in the following \cite{Rosatti2006, Rosatti2008, Rosatti2010, Cozzolino2011}, even if the formal position of the momentum balance appears slightly different.

A strictly closed approach to the last one is that from
\cite{Rosatti2006}, which investigated the numerical computation of
one-dimensional, unsteady, free surface flows over a mobile bed. They
considered a strong interaction between the flow and the erodible
bottom, taking non-conservative terms in the momentum equation into
account and solving the related Riemann problem. The technique was
named AWB (approximately well balanced), and it was applied to schematic and experimental test cases. At the simplest order of approximation, they considered a backward-facing step as the typical scheme concerning the flow across two consecutive computational cells. The force exerted by the step on the fluid is proportional to the height of the step and to a "proper" water depth "in front of" the step.

The word "proper" means that the depth must be measured from the
centroid of the step. The term "in front of" means that the downstream
(or right) depth must be considered if the step is facing backward and
that the upstream (or left) depth must be considered if the step is
facing forward. The force is clearly exerted on the fluid in the flow
direction in the former case, whereas it is opposite to the flow
direction in the latter case. The context of this work is broader than
that of the present work because (as in \cite{Rosatti2008}) a mobile
bed is considered, but the same procedure was also used for a fixed bed elevation \cite{Rosatti2010}.

In particular, \cite{Rosatti2010} thoroughly analysed the Riemann problem for the one-dimensional shallow water equations from theoretical and numerical perspectives. The analysis of the wave at the step leads to a non-conservative crossing of the step in terms of total head. The momentum balance on the step is closed using an integral momentum balance that is very similar to that proposed herein. The only difference is the estimate of the depth on the step, which is performed in \cite{Rosatti2010} using the same technique proposed by \cite{Rosatti2006} and \cite{Bernetti2008}. However, the simple reasoning proposed here does not require a new formulation of the Riemann invariants, and a proper choice of the special depth satisfies both momentum and energy conservation requirements.

A recent contribution to the problem which given by
\cite{Cozzolino2011}, which analysed the shallow water equations on
bottom discontinuities. In this work, the hydrostatic-like pressure
distribution at the step, according the concepts of
\cite{Bernetti2008} and \cite{Rosatti2006}, is also analysed and discussed in the framework of the path-conservative theory of \cite{DLM}. They rebut the preservation of the total head through the bed step \cite{alcrudo01, LeFloch2011, Caleffi2009, Caleffi2015}, claiming that experimental evidence shows that the dissipation on the step is a well-known phenomenon. They consider that the integral momentum balance, taking into account the force that the bed exerts onto the flow, conflicts with the total head conservation hypothesis. In \cite{Cozzolino2011}, they explicitly make reference to the flow detachment and reattachment and eddy recirculation cells, which subtract mechanical energy from the mean flow.

By contrast, the hypothesis supported here is that the dissipation is
evident, but it is not automatically incorporated in the shallow water
scheme. If a proper source term is not added, then the total head is
conserved, and integral momentum conservation can also be preserved by
simply selecting a proper depth on the step.

The authors in \cite{Caleffi2009} proposed a finite volume WENO
scheme, which is fourth-order accurate in space and time, for the
numerical integration of shallow water equations with the bottom slope
source term. The method for managing bed discontinuities is based on a
suitable reconstruction of the conservative variables at the cell
interfaces, coupled with a correction of the numerical flux based on
the local conservation of total energy. Properly selected test cases
show the efficiency of the method in treating discontinuities. 
The method finds a flux correction on bed steps
that takes into account the force exchanged between the step itself
and the fluid, and the flux balancing is achieved not only in still
water conditions but, more generally, also for each steady-state dynamic condition. The analytical treatment of the step that is proposed here is conceived as a deepening of the force estimate on the step itself.

In \cite{Caleffi2015}, using an unified framework consisting of a third-order accurate discontinuous Galerkin scheme, five different numerical methods applied to the free-surface shallow flow simulation on bottom steps are compared.
The role that the treatment of bottom discontinuities plays in the preservation of specific asymptotic conditions is examined. In particular, three widespread approaches based on the motionless steady state balancing \cite{CKL, HSR, PCL} are compared with two approaches \cite{Caleffi2009, Caleffi2015} that are based on the preservation of a moving-water steady state.
The fundamental findings support the concept that the well balancing
of a moving steady state (rather than a well-balanced model in the
case of still water) significantly improves the overall behaviour of the schemes. This perspective is successful both in the framework of the classical finite volume approach  and in the framework of the path-conservative schemes. Finally, in the context of such path-conservative schemes, the fundamental concept of total head conservation on the contact discontinuity over the step provides optimal results in reconstructing well-balanced solutions.

\section{Shallow water equations with source terms: discrepancies between the physics and the scheme}
\label{sec:SWES} 
The present approach is completely based on the necessity of
simultaneously satisfying the SWE as balance laws with source terms on
bed discontinuities and the constraints implicated in such
balance laws, in other words, to respect the schematized physics that we are assuming.

Referring to the unit-width channel, the standard form of SWE without friction is considered: 
\begin{equation}\label{eq:CSWE}
\frac{\partial}{\partial t}\begin{bmatrix}Y\\ q\\z\end{bmatrix}+
\begin{bmatrix}0&1&0\\g\,Y-\frac{q^2}{Y^2}&2\frac{q}{Y}&g\,Y\\0&0&0\end{bmatrix}
\frac{\partial}{\partial x}\begin{bmatrix}Y\\ q\\z\end{bmatrix}
 = \begin{bmatrix}0\\ 0\\0\end{bmatrix};
\end{equation} 

where: $Y(x,t)$, $q(x,t)$ and $z(x,t)$ are the depth, the  specific discharge and the  bottom elevation, respectively; $g$ is the gravity acceleration; $x$ and $t$ are the space and the time, respectively.

A fundamental problem concerning the behaviour of shallow water equations on a bed discontinuity is understanding what physical properties can be conserved and why.
From a purely experimental perspective, it is obvious that a sharp
corner or cusp in the boundary of the flow domain is a source of
vorticity and causes the detachment of eddies of different scales,
which ultimately are the evidence of energy dissipation. There are no
doubts regarding this aspect, which has been proven theoretically and experimentally and can be reproduced using numerical simulations including a turbulence model \cite{Rani2007, Hattori2010}.

Notwithstanding, to understand whether the standard shallow water
scheme can include this aspect is a different problem. In the
classical, well-stated formulation of such a scheme, the vertical
averaging procedure hides some aspects of the physics. The main aspect
that becomes quite artificial is the nature of the dissipation
mechanism, which is shifted from a volume phenomenon to a surface
phenomenon. The standard SWE formulation drops dispersion terms
(depending on the difference from the local velocity and the
vertically averaged velocity) and the diffusion terms due to the eddy
viscosity (if some turbulence model to estimate Reynolds stresses is
supposed to hold). After dropping such terms, the flow over the bottom
is schematized as a piston flow of uniform velocity, which slips on
the bottom itself. At the bottom, a shear stress is exerted, which is
defined as skin friction. In such a scheme, the product of the mean velocity times the shear stress is proportional to the mechanical power dissipated per unit width per unit weight of the liquid. If skin friction is neglected, as here, such dissipated power is zero.

Another mechanism of dissipation in SWE is the hydraulic jump, analogous to the inviscid shock in compressible fluid mechanics. Considering the shallow water framework, the classical hydraulic literature shows that momentum conservation in a jump implies specific energy dissipation. This is intimately due to the structure of the relationships between the flow depth, the total force and the specific energy; see, among others, \cite{Chow, Henderson1966, Chaudry, Valiani2008}.

The interesting conclusion is that in the shallow water scheme, only
two dissipation mechanisms are possible: friction at the bottom
(or at the wetted perimeter for rectangular cross sections that are
not wide) and localized dissipation at jumps and bores.

Another form of energy dissipation in real open channels, which is the
localized dissipation due to singularities in cross section geometry
or bed elevation (bottom steps, cross section widening or
constriction, bridge piers or trash racks), can be taken into account
by adding the classical localized head loss terms, structured as an
empirical dimensionless coefficient times kinetic head; the
coefficient essentially depends on the flow boundary geometry (at
least for sufficiently large Reynolds numbers), the obstacle type and
configuration, and so forth. Specific manuals are addressed for
details \cite{Chow, Henderson1966, Chaudry}. In this context, the only
relevant aspect for a SWE scheme is to stress that the localized head
loss, if significant, must be added to the momentum and/or to the
energy balance equation, and it is not included in the standard formulation.

In this perspective, the main concept presented herein consists of
demonstrating that prescribing the integral momentum balance for a
control volume containing the bed discontinuity, bounded by two cross
sections immediately upstream and downstream of the step, and imposing
a flow depth on the step that is not specifically conceived to
preserve energy for any possible configuration is not a correct
procedure. The word correct refers to the internal consistency of the
model with the physical and mathematical properties of the shallow
water equations. The problem is far from being a formal one because,
as shown in the following, some common choices of the depth on the
step may produce, in some particular cases, a gain in total head in
steady-state flows crossing the step. Being irrefutable that a steady
transition – without energy transfer to the flow by the boundaries or
external work – must not produce energy, the present work intends to
enforce the general idea that, if no extra dissipation term is
inserted in the shallow water formulation, then the energy
conservation principle at bed discontinuities, as proposed by the
general theory of generalized Riemann invariants, holds. However, if a
localized dissipation has to be taken into account, an ad hoc
localized head loss term has to be inserted that depends on an experimental or semi-empirical analysis. In the following, this possibility is not analysed because in principle, adopting the assumption of prescribing a certain specific energy depending on a localized head loss is quite similar to prescribing a certain specific energy imposed by a total head conservation condition.

Two fundamental assumptions are made here. The first is that integral
momentum conservation – in the limits of the fundamental hypothesis of
SWE – must hold. The second is that pressure force must be computed
using the hydrostatic pressure distribution over each vertical, and
this is again for internal consistency with the governing hypothesis of SWE. Moreover, the unit discharge and the total head conservation across the step are assumed. This last couple of assumptions, which are unconditionally assumed in the following, have a larger generality even in unsteady flows, as previously and extensively discussed \cite{alcrudo01, LeFloch2011}.

\section{Some analytical tools}
\label{sec:SAT}
Some definitions and some analytical results are presented here, which
are taken from classical hydraulics and from \cite{Valiani2008}.

Let $Y, \, Y_c$ be the depth and critical depth, $E,
E_c$ be the specific energy and the critical specific energy, and $F,
\, F_c$ be the total force and the critical total force per unit width. The main definitions and their non-dimensional counterparts ($\eta, \, \Gamma, \, \Phi$ as the non-dimensional depth, energy and total force, respectively) are summarized here ($\rho$ is fluid density):

\begin{equation}
         E = Y + \frac{q^{2}}{{2}g\,Y^{2}};
\quad F = \frac{1}{2}\rho g\,Y^{2}  + \frac{\rho\,q^{2}}{Y}
\end{equation} 
\begin{equation}
 Y_c = \sqrt[3]{{\frac{q^{2}}{g}}}; \,\, E_c =\frac{3}{2}Y_c; \,\, F_c = \frac{3}{2}\rho g\,Y_c^2;
          \quad \eta = \frac{Y}{Y_c}; \,\, \Gamma = \frac{E}{E_c}; \,\, \Phi = \frac{F}{F_c}
\end{equation} 
\begin{equation}
         \Gamma = \frac{2}{3}\eta + \frac{1}{3}\left({\frac{1}{\eta}}\right)^{2};
       \quad \Phi = \frac{1}{3}\eta^{2}  + \frac{2}{3}\left({\frac{1}{\eta}}\right)
\end{equation} 

These relationships are extensively discussed in \cite{Valiani2008},
where the role of the Froude number, $\Fr = {U / {\sqrt {g\,Y} }}  =
\eta ^{-3/2}$ ($U$ is the mean velocity of the flow, such that $UY=q$,
assumed as positive everywhere) is also placed into evidence.

Expressions for non-dimensional specific energy and non-dimensional
specific force admit analytical inversion. The solution is summarized
in Tab. \ref{tab:eqtable}, taken from \cite{Valiani2008}, where
$\alpha = \arctan\left({\sqrt{\Gamma _0^3-1}}\right)$ and $\theta  = \arctan\left({\sqrt {\Phi _0^3-1}} \right)$. Physically significant values of $\Gamma _0$ and $\Phi _0$ (both greater than 1) are assumed to be chosen.

\begin{table}
  \begin{center}
  \begin{tabular}{ccc}
  \hline
\rule[-5pt]{0pt}{18pt} Equation & Subcritical solution & Supercritical solution\\
  \hline
\rule[-10pt]{0pt}{30pt} $\Gamma _0  - \frac{2}{3}\eta  - \frac{1}{3}\left( {\frac{1}{\eta }} \right)^2  = 0$ & $\eta_{sb}  = \frac{{\Gamma _0 }}{2}\left[ {1 + 2\cos \left( {\frac{\pi -2 \alpha}{3}} \right)} \right]$ & $\eta_{sp}  = \frac{{\Gamma _0 }}{2}\left[ {1 + 2\cos \left( {\frac{\pi +2 \alpha}{3}} \right)} \right]$ \\
        
\rule[-10pt]{0pt}{25pt} $\Phi _0  - \frac{1}{3}\eta ^2  - \frac{2}{3}\left( {\frac{1}{\eta }} \right) = 0$ &
$\eta_{sb}  = 2\sqrt {\Phi _0 } \;\cos \left( {\frac{\pi - \theta}{3}} \right)$ &
$\eta_{sp}  = 2\sqrt {\Phi _0 } \;\cos \left( {\frac{\pi + \theta}{3}} \right)$ \\
\hline
  \end{tabular}
 \caption{Physically meaningful solutions of equations: $\Gamma = \Gamma_0$ and $\Phi = \Phi_0$}
  \label{tab:eqtable}
   \end{center}
\end{table}

\subsection{Some notes concerning the energy balance}
In the following, the total head conservation on the step  as shown by
\cite{alcrudo01} and \cite{LeFloch2011} is extensively used. If $z_b$
is the bed elevation, this may be written as follows:

\begin{equation} \label{eq:totalhead}
H = z_b + Y + \frac{U^2}{2\, g} = z_b + E = const
\end{equation}

Note that Eq. (\ref{eq:totalhead}) means the conservation of the total mechanical energy per unit weight of the liquid and is commonly used in classical hydraulics as here.

In the recent literature on SWE \cite{Bernetti2008, Fjord2011, Bristeau2015}, energy balances are performed by considering the mechanical energy of the water column that weighs on the unit area (or the same quantity divided by $\rho$):

\begin{equation} \label{eq:toroenergy}
\mathcal{E} = \rho g \, Y\, z_b + \frac{1}{2} \, \rho g\, Y^2 + \frac{1}{2} \, \rho\, U^2\, Y
\end{equation}

The reason is that such energy forms an entropy pair function with a
properly defined entropy flux that can be used to analyse the
properties of the SWE system, in a strict analogy with the Euler
equation in gas dynamics. In \cite{Bernetti2008}, the total energy $T$
of the water included in a volume between two cross sections $x_L$ and
$x_R$ is computed, also taking into account the external work $W$ of the pressure forces at the boundaries of the domain:

\begin{equation}
T=\int_{x_L}^{x_R}\mathcal{E}\,dx\, -\, W
\end{equation}
where $p$ is the liquid pressure, $z$ is the vertical coordinate and
$t$ is time:
\begin{equation}
-W=\int_{z_L}^{z_L+Y_L}p\left(x_R\right)\,U\left(x_R\right)\,dz\,dt\, -\,
\int_{z_R}^{z_R+Y_R}p\left(x_L\right)\,U\left(x_L\right)\,dz\,dt
\end{equation}

By assuming $x_L$ and $x_R$ just upstream and just downstream of the
step and taking the Leibniz rule into account (see
\cite{Bernetti2008} for details), it is possible to compute the rate
of change of the mechanical energy in time as follows:

\begin{eqnarray}
\frac{dT}{dt} & = &
\left(\rho g \, U\, Y\, z_b + \frac{1}{2} \, \rho g\, U\, Y^2 + \frac{1}{2} \, \rho\, U^3\, Y\right)_R + \nonumber\\
& - & \left(\rho g \, U\, Y\, z_b + \frac{1}{2} \, \rho g\, U\, Y^2 + \frac{1}{2} \, \rho\, U^3\, Y\right)_L + \nonumber\\
& + & \left( \frac{1}{2} \, \rho g\, U\, Y^2\right)_R - \left( \frac{1}{2} \, \rho g\, U\, Y^2\right)_L
\end{eqnarray}
that is:
\begin{equation}
\frac{dT}{dt}=\rho g\,q \left[\left(z_b + Y + \frac{U^2}{2\, g}\right)_R-\left(z_b + Y + \frac{U^2}{2\, g}\right)_L\right]
\end{equation}
and finally:
\begin{equation}
\frac{dT}{dt}=\rho g\,q\left(H_R-H_L\right)
\end{equation}

This quantity must be 0 (conservation case) or less than 0
(dissipation case). In general, if friction terms are neglected, then
the inequality holds for inviscid shocks and equality elsewhere
\cite{Bernetti2008, Fjord2011, Bristeau2015, Audusse2015}. As shown here, once $q =
const$ is assumed, as derived from the generalized Riemann invariants
conservation over the step, the general theory concerning the entropy
pair constraints provides (obviously) the same result as classical
hydraulics in stationary flow, which is that the total head cannot
increase in the flow direction. In the following, the total head is
definitively used as a representative of the mechanical energy associated with the flow field.

\section{Subcritical and supercritical flows over bottom discontinuities}\label{sec:SBSPBD}
The computations are organized as follows. A precise value of depth is
supposed to exist as the only one that allows for momentum
conservation. It can be considered as being measured from a bed elevation that is an intermediate value between $z_L$ and $z_R$; the centroid of the step can be the natural reference point, but this is not a necessary assumption in the following because in the total head conservation, only $z_L$ and $z_R$ appear directly, while in the integral momentum balance, only the step height $\left|z_L-z_R\right|$ appears directly.

This value can be expected to be in the range $[\min(Y_L, Y_R),
  \max(Y_L, Y_R)]$, where $Y_L, Y_R$ are the flow depths just upstream
and downstream of the step. The uniqueness of the unit discharge implies that only one between the depths $(Y_L, Y_R)$ can be imposed.
According to the classical theory of flow profiles in open channels,
the upstream $(Y_L)$ depth is imposed when the flow is supercritical,
and the downstream $(Y_R)$ depth is imposed when the flow is
subcritical. The corresponding (having the same total head) downstream
$(Y_R)$ or upstream $(Y_L)$ flow depth is respectively found by simply
considering as prescribed the corresponding specific energy of the
flow ($E_R$ or $E_L$) and finding the other value ($E_L$ or $E_R$) by
the total head conservation condition. The corresponding depth $Y_L$
or $Y_R$ is found directly using the analytical results of
\cite{Valiani2008}, which are summarized in Tab. \ref{tab:eqtable}.

The force on the step is found from the difference between the left
and the right total forces of the stream, where the sign of such force
depends on the step type (backward-facing step (BFS) or forward-facing
step (FFS)). It is easy to verify that this procedure is sufficient
for obtaining a unique value of the depth  {\it on\ } the step, which is the only value of the depth that simultaneously satisfies total head conservation, liquid discharge conservation and integral momentum conservation.

Using the typical nonlinear nature of the depth-energy relationship
and of the depth-total force relationship \cite{Valiani2008}, it is
shown that the depth on the step is included in the expected
range. The behaviour of such depth is computed and described in the
entire range of possible cases and compared with the left depth, right
depth and the corresponding depths computed according to recent models
from the literature, i.e., \cite{Bernetti2008, Rosatti2006, Rosatti2008, Rosatti2010}.

Such models are tested as follows, where $\Delta$ is the step
height. In a subcritical flow, given the prescribed depth $Y_R$
downstream of the step, the corresponding $Y_L$ depth that satisfies the momentum conservation is computed assuming that - as prescribed by such models themselves - the depth on the step is $Y_L - \Delta/2$ (FFS) or $Y_R - \Delta/2$ (BFS). Similar computations are performed, using  $Y_L$ as prescribed, for a supercritical flow.

Moreover, the amount of the total head that is dissipated on the step
$\Delta H = \left(H_L-H_R\right)$ according to such models is
computed, showing that, in some cases, this amount is surprisingly
negative, which means a net gain in the total head. This result is considered proof that the depth on the step must be chosen with proper care.

The specific proposal formulated here is the assumption of total head
conservation, which avoids an unphysical gain of energy, satisfies the constancy of generalized Riemann invariants, and does not introduce fictitious mechanisms of energy dissipation, which are not incorporated in the nature of the SWE scheme.

\begin{figure}
\begin{center}
\includegraphics[width=1.0\textwidth]{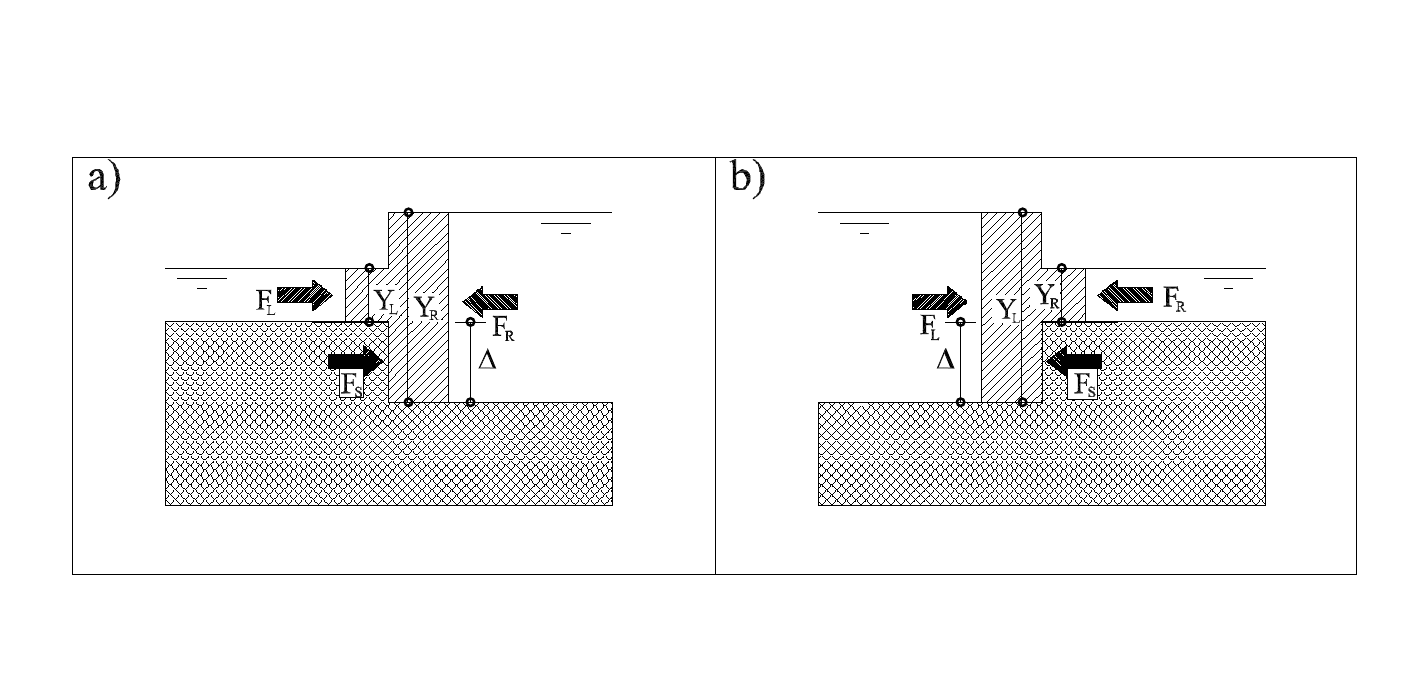}
\end{center}
\caption{Control volume used for the fundamental integral momentum balance. a) Backward-facing step (BFS); b) Forward-facing step (FFS)}
\label{fig:step}
\end{figure}

\subsection{Energy balance on the step}
As a consequence of the previous discussion, the total head
conservation on the step is assumed such that the specific energy
balance can be written  for both the backward-facing step (BFS,
Fig. 1a) and the forward-facing step (FFS, Fig. 1b). This balance may
be written as follows:
\begin{equation} \label{eq:enstep}
\textrm{BFS:}\,\,E_L+\Delta=E_R\,;\quad \textrm{FFS:}\,\,E_L-\Delta=E_R
\end{equation}

In the following, repeating the acronyms BFS and FFS is avoided for
brevity, and the symbol $\pm$ is used, where $+$ refers to the BFS and
$-$ refers to the FFS; thus, the above energy balance is: 

\begin{equation} \label{eq:enstepext}
\left(Y_L+\frac{q^2}{2\,g\,Y_L^2}\right)\pm\Delta=
\left(Y_R+\frac{q^2}{2\,g\,Y_R^2}\right)
\end{equation}

Using the equality $q^2=g\,Y_c^3$,  taking into account that
$\Delta/E_c=({2}/{3})\,\delta$ (with $\delta=\Delta/Y_c$) and forming
non-dimensional Eq. (\ref{eq:enstep})  and Eq. (\ref{eq:enstepext}) by
dividing each term for $E_c$, the non-dimensional energy balance over
the step (in compact and extended forms, respectively) is:

\begin{equation} \label{eq:enstepnd}
\Gamma_L \pm \frac{2}{3}\,\delta = \Gamma_R
\end{equation}
\begin{equation} \label{eq:enstepndext}
\frac{2}{3}\,\eta_L+\frac{1}{3}\,\frac{1}{\eta_L^2} \pm \frac{2}{3}\,\delta =
\frac{2}{3}\,\eta_R+\frac{1}{3}\,\frac{1}{\eta_R^2}
\end{equation}

This conservation law is not dependent on the specific definition of
the flow depth on the step, and it is considered to be valid based on the previously discussed literature.

\subsection{Momentum balance on the step}
The integral momentum balance is written for both the BFS and the
FFS. The total force is considered to be positive if directed
according to the flow direction (see Fig. \ref{fig:step}). This
balance, where $F_L$ is the total force upstream of the step, $F_R$ is
the total force downstream of the step, and $F_S$ is the total force
on the step, may be written as follows:

\begin{equation} \label{eq:intmomstep}
\textrm{BFS:}\,\,F_L+F_S=F_R\,;\quad \textrm{FFS:}\,\,F_L-F_S=F_R
\end{equation}

In extended form,  Eq. (\ref{eq:intmomstep}) reads as follows: 
\begin{equation} \label{eq:intmomstepext}
\left(\frac{1}{2}\,\rho\,g\,Y_L^2+\rho\frac{q^2}{Y_L}\right)\pm\rho\,g\,Y_S\,\Delta=
\left(\frac{1}{2}\,\rho\,g\,Y_R^2+\rho\frac{q^2}{Y_R}\right)
\end{equation}

In Eq. (\ref{eq:intmomstepext}), $Y_S$ is the flow depth $\textit{on}$
the step; being $q^2=g\,Y_c^3$,  taking into account that
$\rho\,g\,Y_S\,\Delta/F_c=({2}/{3})\,\eta_S\,\delta$ ($\eta_S$ and
$\delta$ are the non-dimensional depth on the step and the
non-dimensional height of the step, respectively) and forming
non-dimensional Eq. (\ref{eq:intmomstep}) and
Eq. (\ref{eq:intmomstepext})  by dividing each term for $F_c$, the
non-dimensional momentum balance over the step (in compact and
extended forms, respectively) is:

\begin{equation} \label{eq:intmomstepnd}
\Phi_L \pm \frac{2}{3}\,\eta_S\,\delta = \Phi_R
\end{equation}
\begin{equation} \label{eq:intmomstepndext}
\frac{1}{3}\,\eta_L^2+\frac{2}{3}\,\frac{1}{\eta_L} \pm \frac{2}{3}\,\eta_S\,\delta =
\frac{1}{3}\,\eta_R^2+\frac{2}{3}\,\frac{1}{\eta_R}
\end{equation}

Note that the total head conservation (\ref{eq:enstepndext}) is not
influenced by the specific definition of $Y_S$ (and $\eta_S$),
which appear only in the momentum balance. Finally, the only depth on the step that satisfies the integral momentum balance is:

\begin{equation} \label{eq:eta_S}
\eta_S=\pm\frac{\Phi_R-\Phi_L}{\left(2/3\right)\,\delta}
\end{equation}
where the sign $+$ is valid for BFS and the sign $-$ is valid for the FFS.

This value is shown to satisfy the condition $\eta_S\in\left[\min\left(\eta_L , \eta_R\right); \max\left(\eta_L , \eta_R\right)\right]$ (see appendix A), which can also be reasonably expected on the basis of physical reasoning.

In contrast, different estimates of the force over the step and the
corresponding depth on the step ($Y_S=Y_S^{RF}$) are given following
the method of \cite{Bernetti2008, Rosatti2006, Rosatti2008,
  Rosatti2010, Cozzolino2011}. The acronym RF is used in the following
notations simply because \cite{Rosatti2006} was the first in chronological order. All these methods are based on the same criterion, which links $Y_S^{RF}$ to $Y_L$ for FFS and $Y_S^{RF}$ to $Y_R$ for BFS. This can be summarized by considering the balance Eq. (\ref{eq:intmomstepext}) with:
\begin{equation} \label{eq:star}
\textrm{BFS:} \,\, Y_S^{RF}=Y_R-\Delta/2; \quad \textrm{FFS:} \,\, Y_S^{RF}=Y_L-\Delta/2
\end{equation}

\subsection{Subcritical flow on a forward-facing step (SBFFS)}
\begin{figure}
\begin{center}
\includegraphics[width=1.0\textwidth]{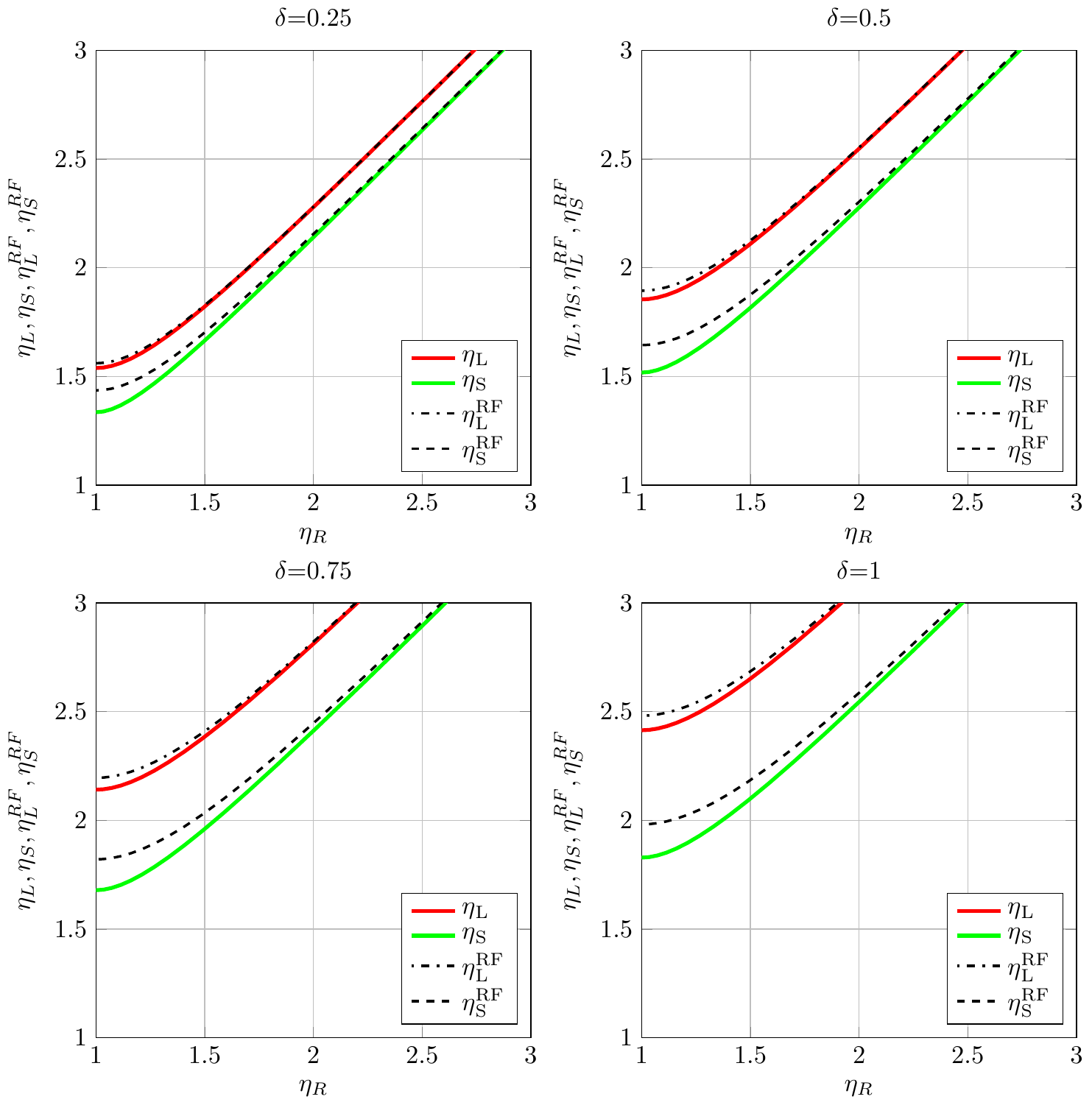}
\end{center}
\caption{SBFFS: left and step non-dimensional depths {\vs} right
  non-dimensional depth for different values of the non-dimensional
  step height according to the present method and the RF \cite{Rosatti2006} and similar methods}
\label{fig:SBFFS1}
\end{figure}

The right condition is assumed as prescribed, following classical hydraulics  for subcritical flows ($\Fr_R < 1$):
($Y_R,\,E_R,\,F_R;\,\eta_R,\,\Gamma_R,\,\Phi_R$) are given. From
Eq. (\ref{eq:enstepndext}), choosing the $-$ sign, the left
non-dimensional depth can be found, using the results summarized in Tab. \ref{tab:eqtable}, $\eta_L=\eta_{sb}$ as unknown and $\Gamma_{0}=\Gamma_R+\frac{2}{3}\,\delta$.
Once the couple of depths $\left(\eta_L,\,\eta_R\right)$ are known, the depth on the step $\eta_S$ can be obtained from Eq. (\ref{eq:eta_S}), choosing the $-$ sign. This value, which satisfies the condition $\eta_R \leq \eta_S \leq \eta_L$, as shown in \ref{sec:app_A}, is plotted in Fig. (\ref{fig:SBFFS1}).

\begin{figure}
\begin{center}
\includegraphics[width=1.0\textwidth]{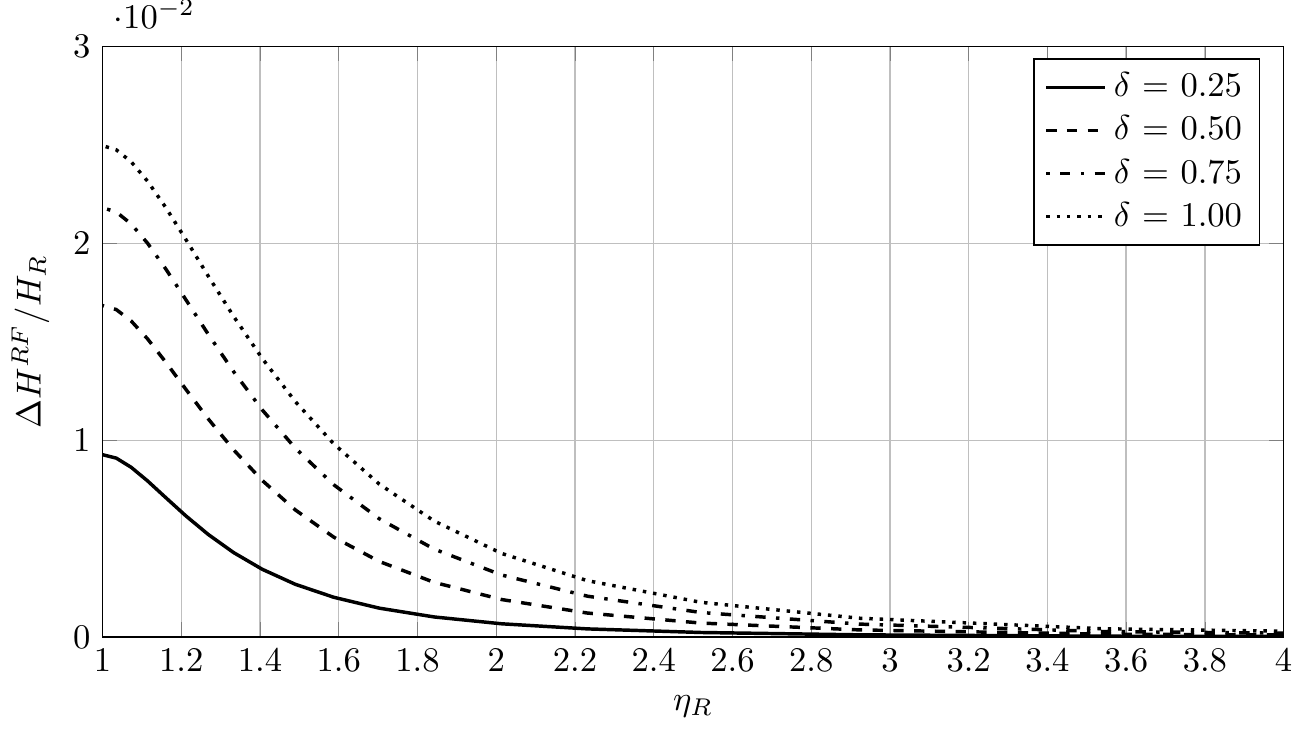}
\end{center}
\caption{SBFFS: relative energy dissipation at the step according to RF \cite{Rosatti2006} and similar methods for different values of the non-dimensional step height}
\label{fig:SBFFS2}
\end{figure}

The left non-dimensional depth that satisfies the momentum balance
according to \cite{Bernetti2008, Rosatti2006, Rosatti2010, Cozzolino2011} is $\eta_L^{RF}$, and the corresponding non-dimensional depth on the step is $\eta_S^{RF}=\eta_L^{RF} - \delta/2$. This value is found from the non-dimensional momentum balance:

\begin{equation}
\frac{1}{3}\,\ \left(\eta_L^{RF}\right)^{2}+\frac{2}{3}\,\frac{1}{\eta_L^{RF}}-\frac{2}{3}\,\delta\,\left(\eta_L^{RF}-\frac{1}{2}\,\delta\right) =\Phi_R
\end{equation}

The subcritical solution of this (3rd degree algebraic) equation is
$\eta_L^{RF}$. This value and the value of $\eta_S^{RF}$ are also
plotted in Fig. (\ref{fig:SBFFS1}). Significant differences from the
method proposed herein can be observed just near the critical state, whereas negligible differences occur for low right Froude numbers, as can be expected on the basis of physical reasoning.
The corresponding non-dimensional specific energy $\Gamma_L^{RF}$ can
be easily computed, which allows the relative total head dissipation
on the step to be computed according to \cite{Bernetti2008, Rosatti2006, Rosatti2010, Cozzolino2011}: 

\begin{equation}
\frac{\Delta H^{RF}}{H_R}=\frac{H_L^{RF}-H_R}{H_R}=\frac{\Gamma_L^{{RF}}-
\left[\Gamma_R+\left(2/3\right)\,\delta\right]}{\left[\Gamma_R+\left(2/3\right)\,\delta\right]}
\end{equation}

This is plotted in Fig. (\ref{fig:SBFFS2}) and shows a relative total
head dissipation on the order of $1 \cdot 10^{-2}$ for non-dimensional
step heights of less than $1$. As a simplified conclusion, the energy
conservation criterion and RF criterion both appear to provide
reasonable results for the SBFFS, even though they are highly different  from a theoretical perspective.

\subsection{Subcritical flow on a backward-facing step without choking (SBBFS)}
\begin{figure}
\begin{center}
\includegraphics[width=1.0\textwidth]{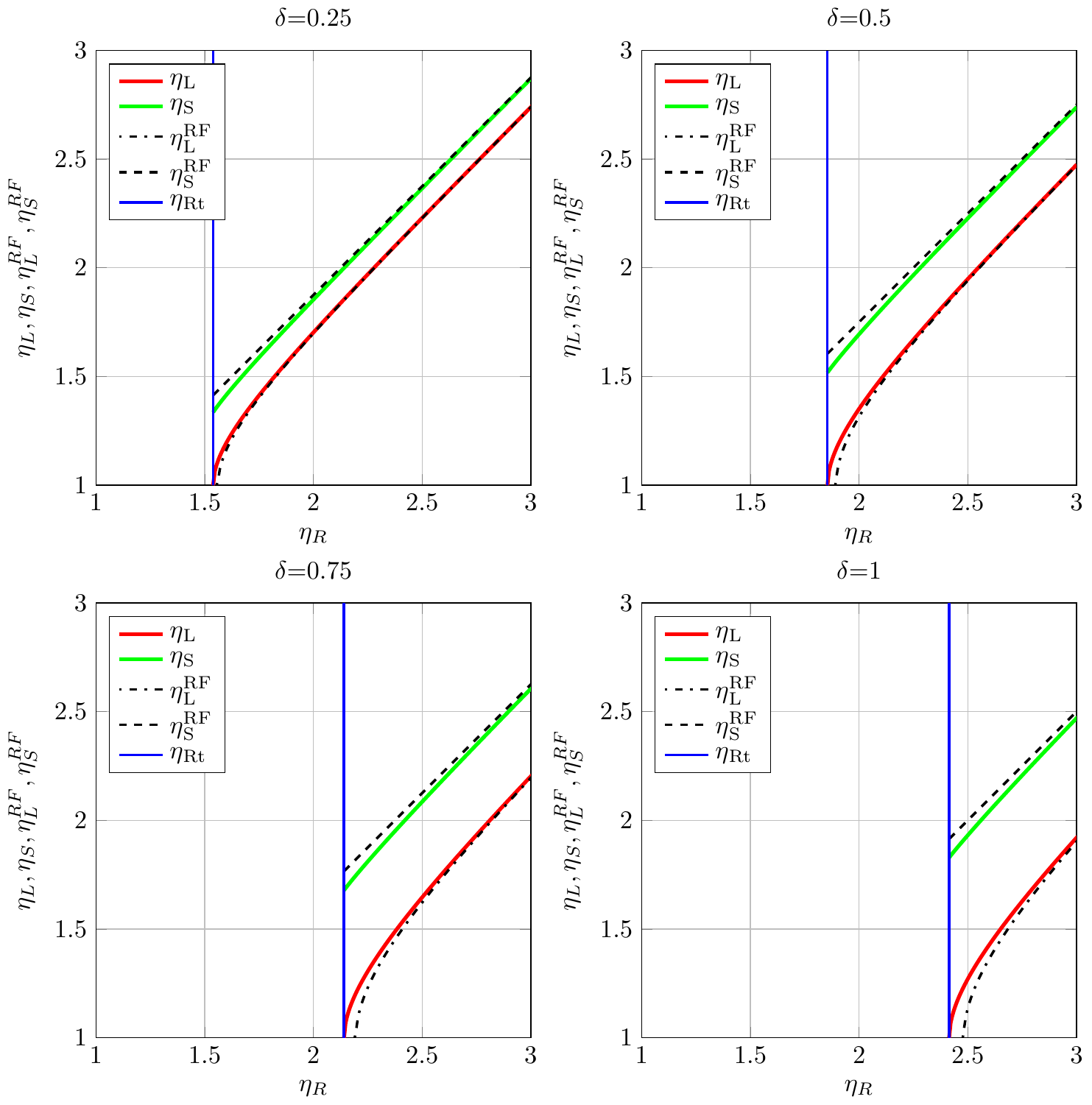}
\end{center}
\caption{SBBFS: left and step non-dimensional depths {\vs} right non-dimensional depth for different values of the non-dimensional step height according the present method and the RF \cite{Rosatti2006} and similar methods}
\label{fig:SBBFS1}
\end{figure}

The right condition is assumed as prescribed, following classical hydraulics  for subcritical flows ($\Fr_R < 1$):
($Y_R,\,E_R,\,F_R;\,\eta_R,\,\Gamma_R,\,\Phi_R$) are given. From
Eq. (\ref{eq:enstepndext}), choosing the $+$ sign, the left
non-dimensional depth can be found, using the results summarized in Tab. \ref{tab:eqtable}, $\eta_L=\eta_{sb}$ as unknown and $\Gamma_{0}=\Gamma_R-\frac{2}{3}\,\delta$.
The necessary condition for obtaining a couple of subcritical solutions, which is to avoid flow choking (i.e., to obtain $\Gamma_0 \ge 1$), is the following:

\begin{equation}
\Gamma_R\geq 1+\frac{2}{3}\,\delta
\end{equation}
which defines a threshold value $\eta_{Rt}$ of the right non-dimensional depth: under this value, flow blocking occurs.
Once the couple of depths $\left(\eta_L,\,\eta_R\right)$ are known, the depth on the step, $\eta_S$, can be obtained from Eq. (\ref{eq:eta_S}), choosing the $+$ sign. This value, which satisfies the condition $\eta_L \leq \eta_S \leq \eta_R$, is plotted in Fig. (\ref{fig:SBBFS1}). 

\begin{figure}
\begin{center}
\includegraphics[width=1.0\textwidth]{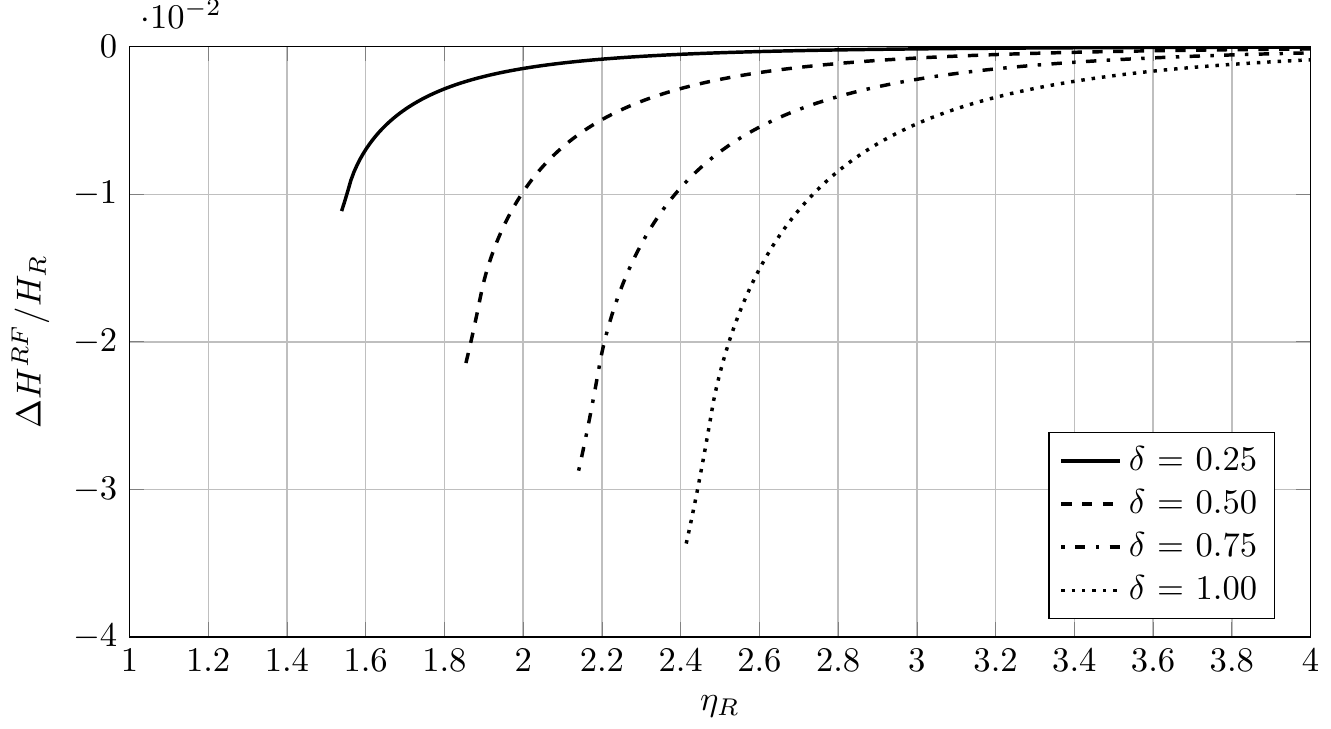}
\end{center}
\caption{SBBFS: relative energy dissipation at the step according RF \cite{Rosatti2006} and similar methods for different values of the non-dimensional step height}
\label{fig:SBBFS2}
\end{figure}

According to \cite{Bernetti2008, Rosatti2006, Rosatti2010,
  Cozzolino2011}, the non-dimensional depth on the step is
$\eta_S^{RF}=\eta_R - \delta/2$, whereas the left non-dimensional
depth that satisfies the momentum balance according to \cite{Bernetti2008, Rosatti2006, Rosatti2010, Cozzolino2011} is $\eta_L^{RF}$; this value is found from the non-dimensional momentum balance:

\begin{equation}
\frac{1}{3}\,\left(\eta_L^{RF}\right)^2+\frac{2}{3}\,\frac{1}{\eta_L^{RF}}=\Phi_R-\frac{2}{3}\,\delta\,\left(\eta_R-\frac{1}{2}\,\delta\right)
\end{equation}

The subcritical solution of this (3rd degree algebraic) equation is
$\eta_L^{RF}$. This value and the value of $\eta_S^{RF}$ are also
plotted in Fig. (\ref{fig:SBBFS1}). Significant differences from the
method proposed herein can be observed just near the threshold state and for the largest values of non-dimensional step heights, as can be expected on the basis of physical reasoning.
The corresponding non-dimensional specific energy $\Gamma_L^{RF}$ can
be easily computed, which allows the relative total head dissipation
on the step to be computed according to \cite{Bernetti2008, Rosatti2006, Rosatti2010, Cozzolino2011}: 

\begin{equation}
\frac{\Delta H^{RF}}{H_R}=\frac{H_L^{RF}-H_R}{H_R}=
\frac{\left[\left(2/3\right)\,\delta+\Gamma_L^{RF}\right]-\Gamma_R}{\Gamma_R}
\end{equation}

This is plotted in Fig. (\ref{fig:SBBFS2}) and shows a negative
relative total head dissipation on the order of $1 \cdot 10^{-2}$ for
non-dimensional step heights of less than $1$. This result means that
a net gain of total head occurs in this case, which is clearly a
physical paradox. From a practical perspective, it is recommended to
avoid the RF method for the SBBFS case without choking to avoid an unphysical gain of energy crossing the step.

\subsection{Backward-facing step with choking (BFSCHK)}
\begin{figure}
\begin{center}
\includegraphics[width=1.0\textwidth]{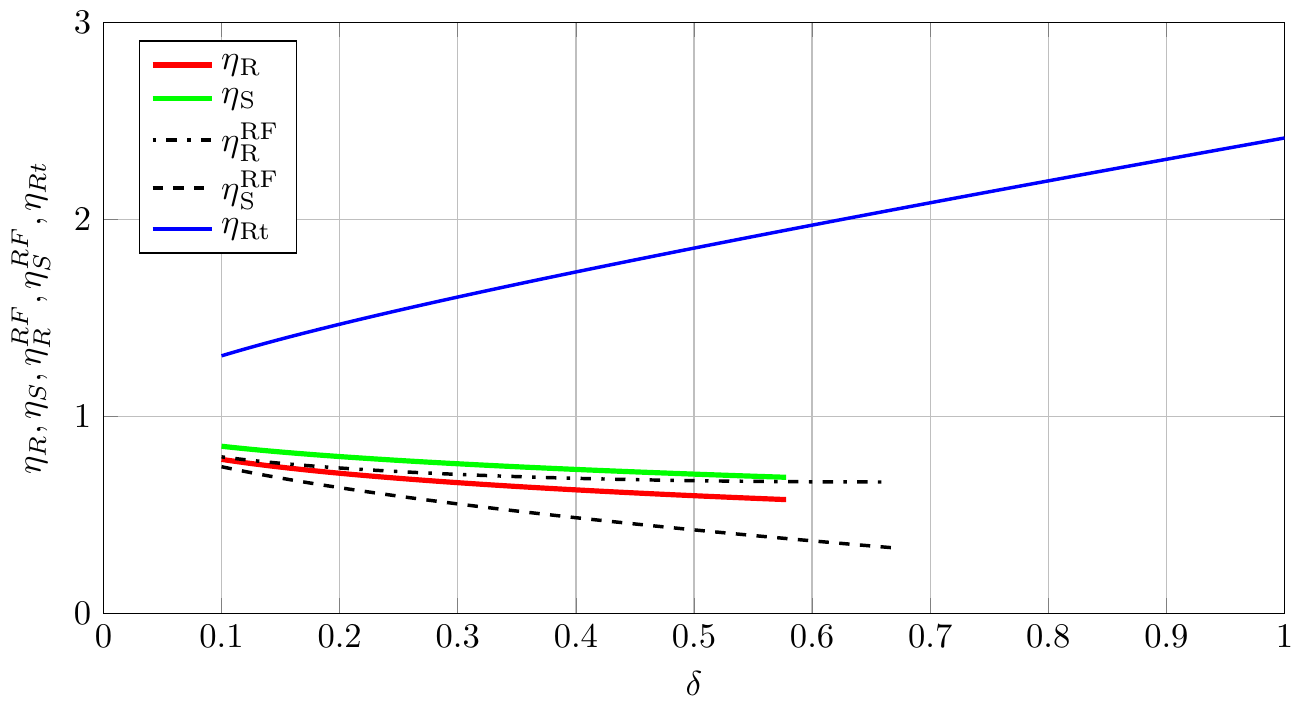}
\end{center}
\caption{BFSCHK: right and step non-dimensional depths {\vs} the
  non-dimensional step height in the case of choking. Lines are interrupted to consider fully submerged step only}
\label{fig:BFSCHK1}
\end{figure}

Flow choking occurs when an attempt is made to prescribe the right
subcritical flow conditions and when the right depth is not
sufficiently large (or similarly, the right Froude number is not
sufficiently low or the right specific energy of the flow is not
sufficiently large), more specifically, when
\begin{equation} \label{eq:thresholdBFS}
\Gamma_R<1+\frac{2}{3}\,\delta
\end{equation}
In this case, the left state just upstream of the step shifts to the
critical state, and the right state just downstream of the step shifts
to the only supercritical state that has the minimum specific energy that is necessary and sufficient to pass the step:
\begin{equation} \label{eq:leftcrit}
\Gamma_L=1;\,\,\eta_L=1;\,\,\Phi_L=1
\end{equation}
\begin{equation}
\Gamma_R=1+\frac{2}{3}\,\delta\,\,\Rightarrow\eta_{R}
\end{equation}
The right supercritical depth is found using the inversion method, which is summarized in Tab. \ref{tab:eqtable}.
Once the couple of depths $\left(\eta_L=1,\,\eta_R\right)$ are known, the depth on the step, $\eta_S$, can be obtained from Eq. \eqref{eq:eta_S}, choosing the $+$ sign and $\Phi_L=1$. This value, which satisfies the condition $\eta_R \leq \eta_S \leq \eta_L=1$, is plotted in Fig. (\ref{fig:BFSCHK1}) as a function of the non-dimensional height of the step. In the same figure, the threshold value of the non-dimensional right depth $\eta_{Rt}$, as defined by Eq. (\ref{eq:thresholdBFS}), is plotted. If the prescribed right non-dimensional depth is less than the threshold value, then flow choking occurs and the critical condition occurs at the left side.

\begin{figure}
\begin{center}
\includegraphics[width=1.0\textwidth]{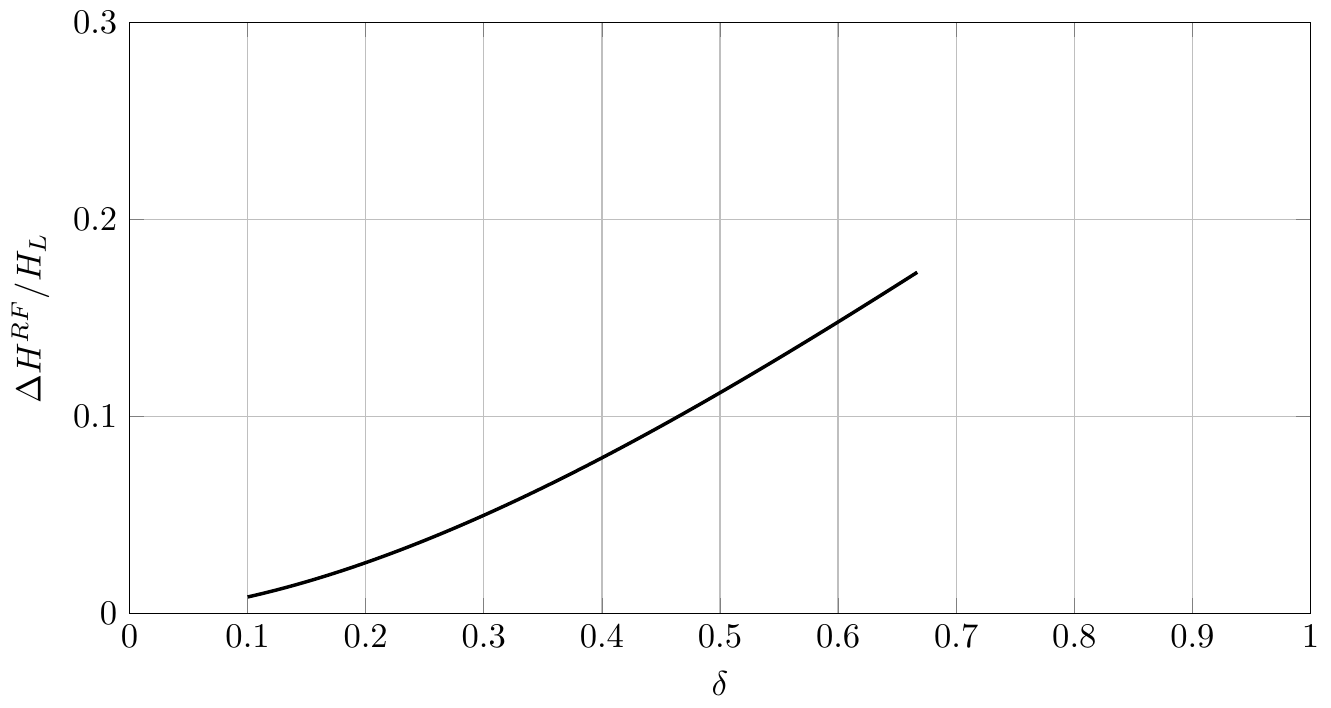}
\end{center}
\caption{BFSCHK: relative energy dissipation at the step according to RF \cite{Rosatti2006} and similar methods {\vs} the non-dimensional step height. The line is interrupted to consider fully submerged step only}
\label{fig:BFSCHKC2}
\end{figure}

The right non-dimensional depth that satisfies the momentum balance
according to \cite{Bernetti2008, Rosatti2006, Rosatti2010,
  Cozzolino2011} is $\eta_R^{RF}$, and the corresponding
non-dimensional depth on the step is $\eta_S^{RF}=\eta_R^{RF} -
\delta/2$. Because the left state is the critical one, the value of  $\eta_R^{RF}$ is found from the non-dimensional momentum balance:

\begin{equation}
1=\frac{1}{3}\,\left(\eta_R^{RF}\right)^2+\frac{2}{3}\,\frac{1}{\eta_R^{RF}}-\frac{2}{3}\,\delta\,\left(\eta_R^{RF}-\frac{1}{2}\,\delta\right)
\end{equation}

The supercritical solution of this (3rd degree algebraic) equation is
$\eta_R^{RF}$. This value and the value of $\eta_S^{RF}$ are also
plotted in Fig. (\ref{fig:BFSCHK1}) as a function of the
non-dimensional height of the step in the range of a fully submerged
step. In this case, significant differences from the method proposed herein can be observed in the entire range of $\delta$.
The corresponding non-dimensional specific energy $\Gamma_R^{RF}$ can
be easily computed, which allows the relative total head dissipation
on the step to be computed according to \cite{Bernetti2008, Rosatti2006, Rosatti2010, Cozzolino2011}: 

\begin{equation}
\frac{\Delta H^{RF}}{H_L}=\frac{H_L-H_R^{RF}}{H_L}=
\frac{\left[\left(2/3\right)\,\delta+1\right]-\Gamma_R^{RF}}{\left[\left(2/3\right)\,\delta+1\right]}
\end{equation}

This is plotted in Fig. (\ref{fig:BFSCHKC2}) and shows a relative
total head dissipation on the order of $1 \cdot 10^{-1}$ for
non-dimensional step heights less than approximately $0.7$. Such a
dissipation that is a strongly increasing function of the step height
$\delta$ is a significantly different result from the conservative
method proposed herein.  

\subsection{Supercritical flow on a backward-facing step (SPBFS)}
\begin{figure}
\begin{center}
\includegraphics[width=1.0\textwidth]{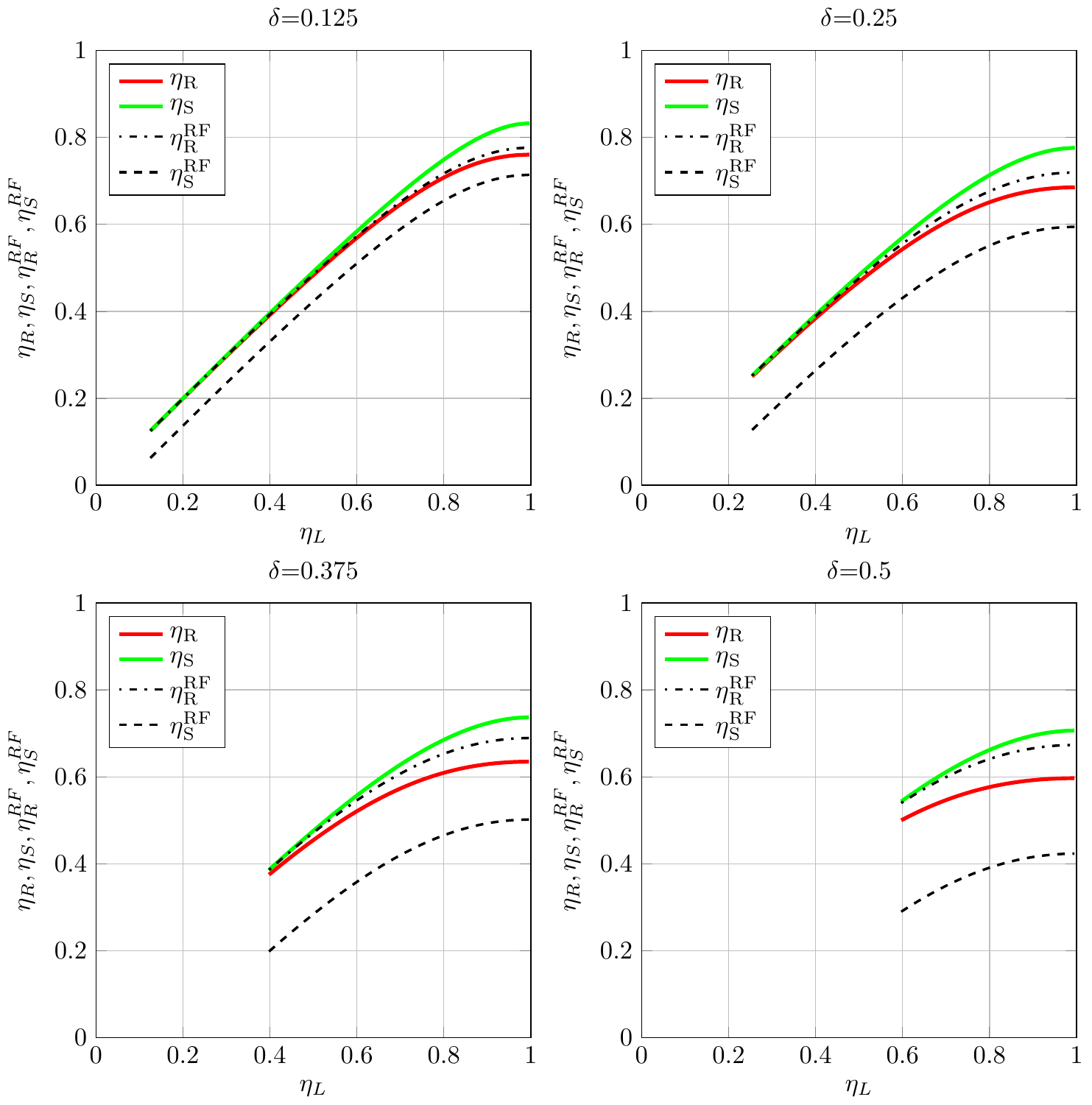}
\end{center}
\caption{SPBFS: right and step non-dimensional depths {\vs} left
  non-dimensional depth according to the present method and the RF \cite{Rosatti2006} and similar methods for different values of the non-dimensional step height. Lines are interrupted to consider fully submerged step only}
\label{fig:SPBFS1}
\end{figure}

The left condition is assumed as prescribed, following classical hydraulics  for supercritical flows ($\Fr_L > 1$):
($Y_L,\,E_L,\,F_L;\,\eta_L,\,\Gamma_L,\,\Phi_L$) are given. From
Eq. \eqref{eq:enstepndext}, choosing the $+$ sign, the right
non-dimensional depth can be found, using the results summarized in Tab. \ref{tab:eqtable}, $\eta_R=\eta_{sp}$ as unknown and $\Gamma_{0}=\Gamma_L+\frac{2}{3}\,\delta$.
Once the couple of depths $\left(\eta_L,\,\eta_R\right)$ are known, the depth on the step, $\eta_S$, can be obtained from Eq. \eqref{eq:eta_S}, choosing the $+$ sign. This value, which satisfies the condition $\eta_R \leq \eta_S \leq \eta_L$, is plotted in Fig. (\ref{fig:SPBFS1}).

\begin{figure}
\begin{center}
\includegraphics[width=1.0\textwidth]{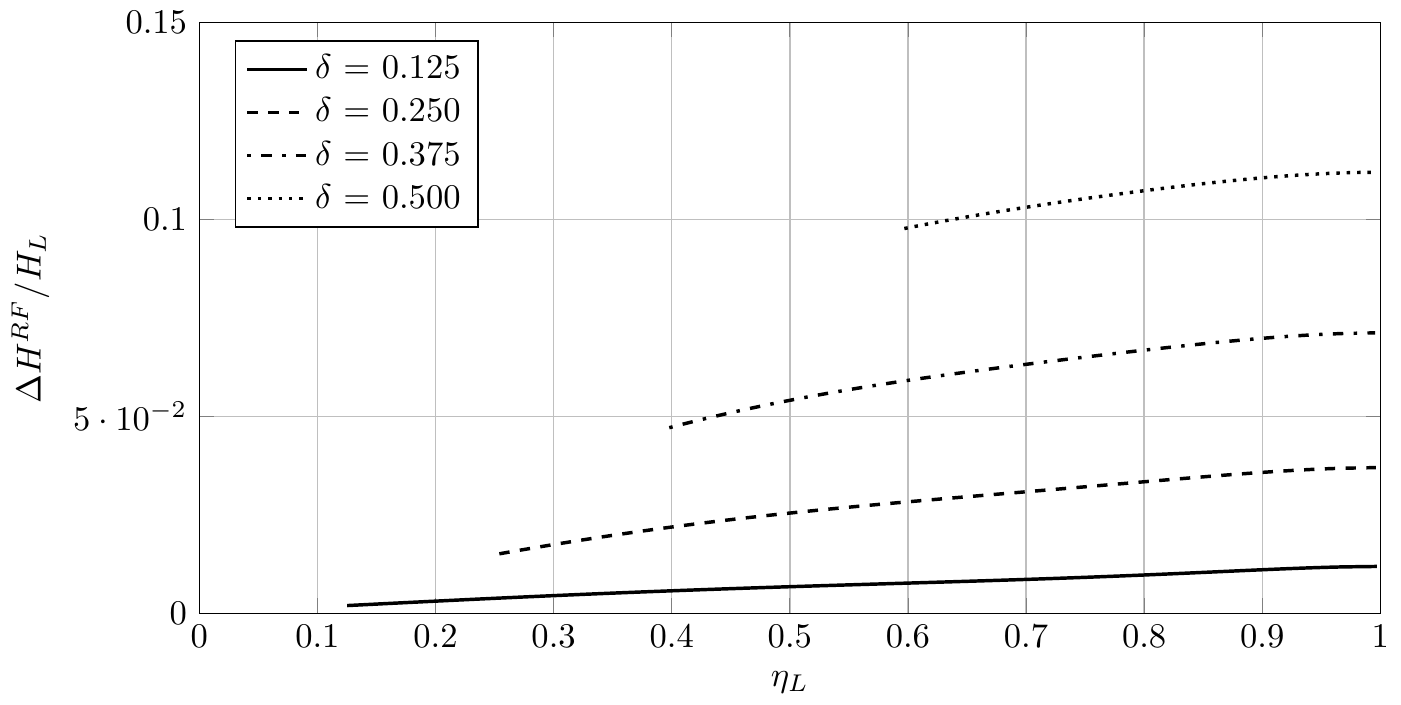}
\end{center}
\caption{SPBFS: relative energy dissipation at the step according to RF \cite{Rosatti2006} and similar methods for different values of the non-dimensional step height.
Lines are interrupted to consider fully submerged step only.}
\label{fig:SPBFS2}
\end{figure}

The right non-dimensional depth that satisfies the momentum balance
according to \cite{Bernetti2008, Rosatti2006, Rosatti2010, Cozzolino2011} is $\eta_R^{RF}$, and the corresponding non-dimensional depth on the step is $\eta_S^{RF}=\eta_R^{RF} - \delta/2$. This value is found from the non-dimensional momentum balance:

\begin{equation}
\Phi_L=\frac{1}{3}\,\left(\eta_R^{RF}\right)^2+\frac{2}{3}\,\frac{1}{\eta_R^{RF}}\,-\frac{2}{3}\,\delta\,\left(\eta_R^{RF}-\frac{1}{2}\,\delta\right)
\end{equation}

The supercritical solution of this (3rd degree algebraic) equation is
$\eta_R^{RF}$. This value and the value of $\eta_S^{RF}$ are also
plotted in Fig. (\ref{fig:SPBFS1}). Significant differences from the
method proposed herein can be observed just near the critical state, whereas negligible differences occur for high right Froude numbers, as expected on the basis of physical reasoning.
The corresponding non-dimensional specific energy $\Gamma_R^{RF}$ can
be easily computed, which allows the relative total head dissipation
on the step to be computed according to \cite{Bernetti2008, Rosatti2006, Rosatti2010, Cozzolino2011}:

\begin{equation}
\frac{\Delta H^{RF}}{H_L}=\frac{H_L-H_R^{RF}}{H_L}=
\frac{\left[\left(2/3\right)\,\delta+\Gamma_L\right]-\Gamma_R^{{RF}}}{\left[\left(2/3\right)\,\delta+\Gamma_L\right]}
\end{equation}

This is plotted in Fig. (\ref{fig:SPBFS2}) and shows a relative total
head dissipation up to the order of $1 \cdot 10^{-1}$ for
non-dimensional step heights of less than $0.5$. Moreover, in this
case, there is a significantly different result from the conservative
method proposed herein. A general comment can be quite similar to the
BFS case with choking, as can be expected because both the left and
right depths are supercritical. 

\subsection{Supercritical flow on a forward-facing step without choking (SPFFS)}
\begin{figure}
\begin{center}
\includegraphics[width=1.0\textwidth]{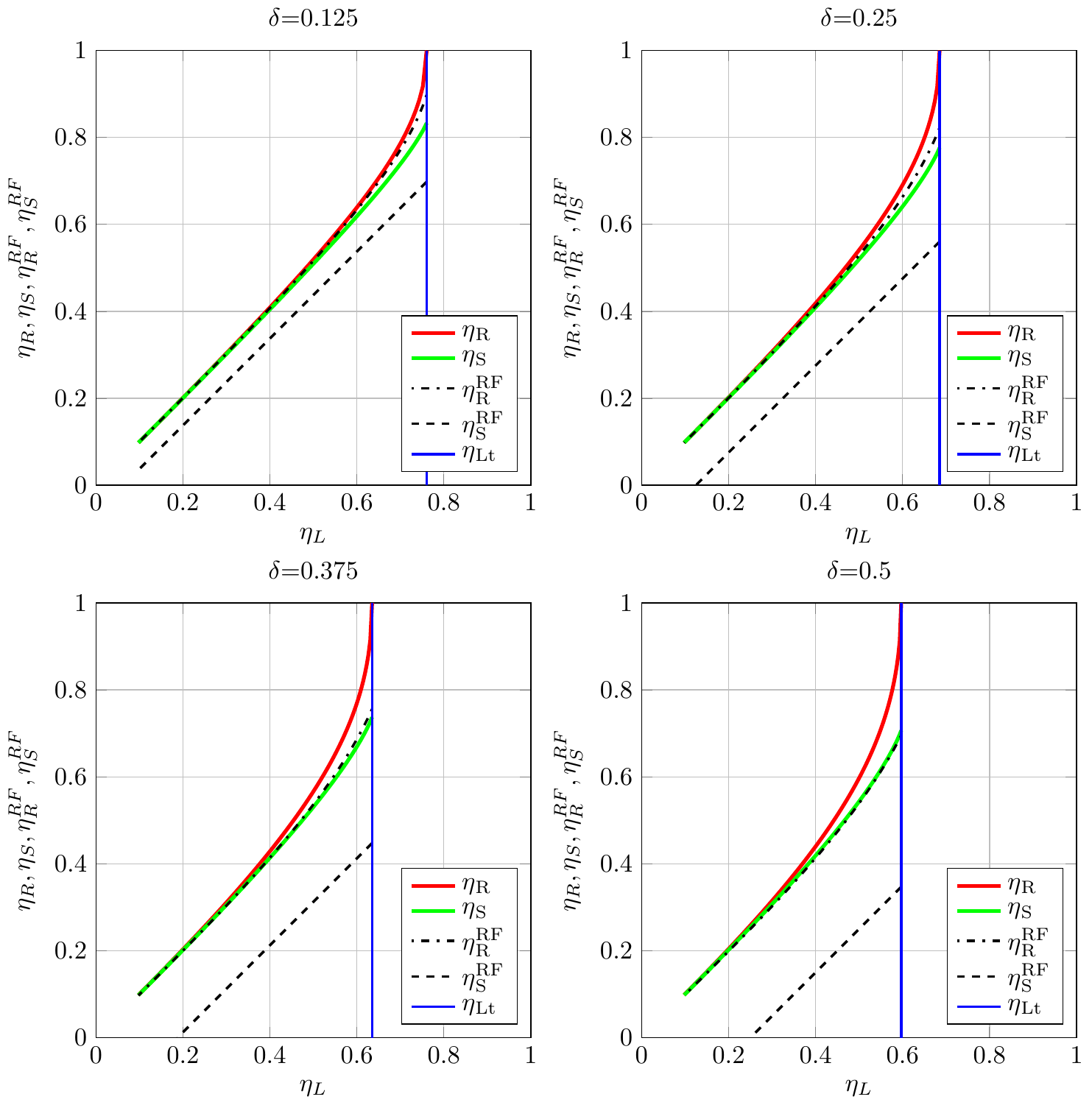}
\end{center}
\caption{SPFFS: right and step non-dimensional depths {\vs} left
  non-dimensional depth according to the present method and the RF
  \cite{Rosatti2006} and similar methods for different values of the
  non-dimensional step height. Only fully submerged steps are considered}
\label{fig:SPFFS1}
\end{figure}

The left condition is assumed as prescribed, following classical hydraulics  for subcritical flows ($\Fr_L > 1$):
($Y_L,\,E_L,\,F_L;\,\eta_L,\,\Gamma_L,\,\Phi_L$) are given. From
Eq. \eqref{eq:enstepndext}, choosing the $-$ sign, the right
non-dimensional depth can be found, using the results summarized in  Tab. \ref{tab:eqtable}, $\eta_R=\eta_{sp}$ as unknown and $\Gamma_{0}=\Gamma_L-\frac{2}{3}\,\delta$.
The necessary condition to obtain a couple of supercritical solutions, which is to avoid flow choking (i.e., to obtain $\Gamma_0 \ge 1$), is the following:
\begin{equation}
\Gamma_L\geq 1+\frac{2}{3}\,\delta
\end{equation}
which defines a threshold value $\eta_{Lt}$ of the left non-dimensional depth: over this value, flow blocking occurs.
Once the couple of depths $\left(\eta_L,\,\eta_R\right)$ are known, the depth on the step, $\eta_S$, can be obtained from Eq. \eqref{eq:eta_S}, choosing the $-$ sign. This value, which satisfies the condition $\eta_L \leq \eta_S \leq \eta_R$, is plotted in Fig. (\ref{fig:SPFFS1}).

\begin{figure}
\begin{center}
\includegraphics[width=1.0\textwidth]{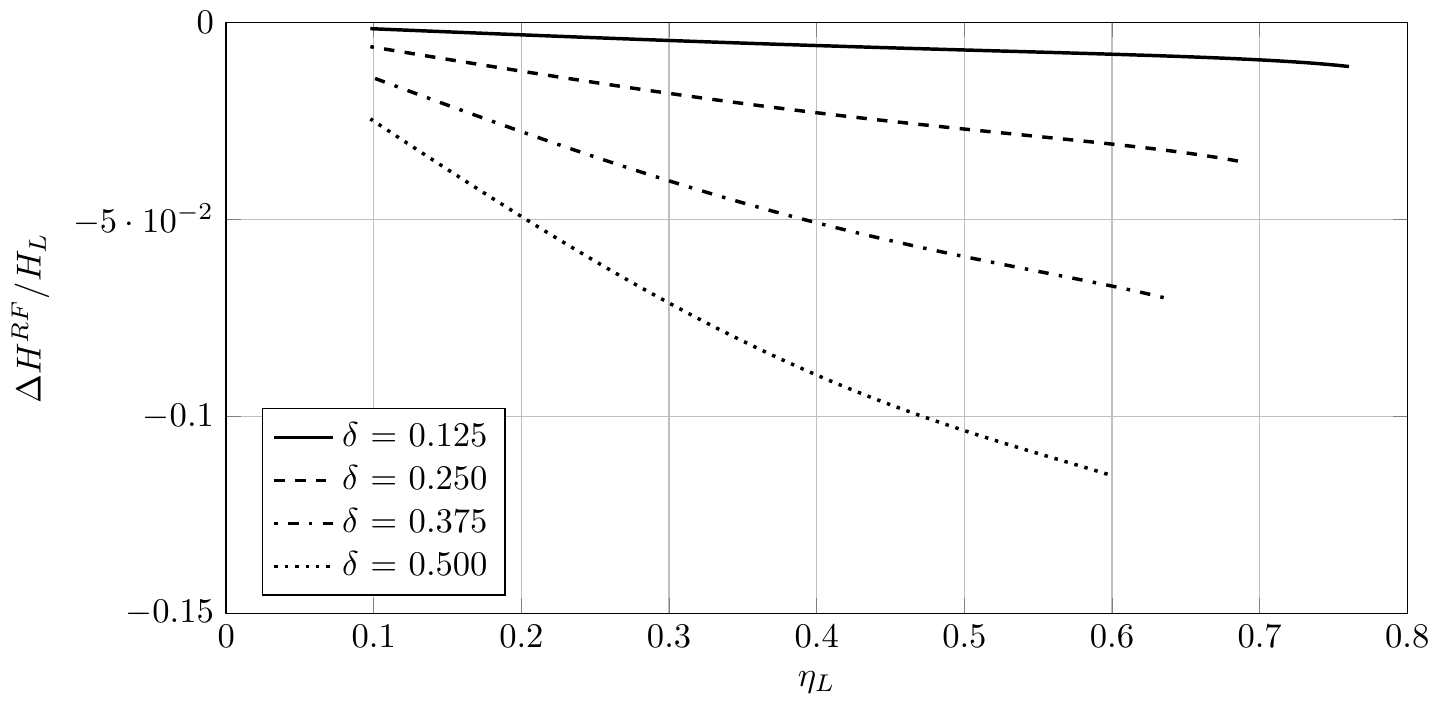}
\end{center}
\caption{SPFFS: relative energy dissipation at the step according to
  RF \cite{Rosatti2006} and similar methods for different values of
  the non-dimensional step height. Only fully submerged steps are considered}
\label{fig:SPFFS2}
\end{figure}

According to \cite{Bernetti2008, Rosatti2006, Rosatti2010,
  Cozzolino2011}, the non-dimensional depth on the step is
$\eta_S^{RF}=\eta_L - \delta/2$, whereas the left non-dimensional
depth that satisfies the momentum balance according to \cite{Bernetti2008, Rosatti2006, Rosatti2010, Cozzolino2011} is $\eta_R^{RF}$; this value is found from the non-dimensional momentum balance:

\begin{equation}
\frac{1}{3}\,\left(\eta_R^{RF}\right)^2+\frac{2}{3}\,\frac{1}{\eta_R^{RF}}=\Phi_L-\frac{2}{3}\,\delta\,\left(\eta_L-\frac{1}{2}\,\delta\right)
\end{equation}

The supercritical solution of this (3rd degree algebraic) equation is
$\eta_R^{RF}$. This value and the value of $\eta_S^{RF}$ are also
plotted in Fig. (\ref{fig:SPFFS1}). Significant differences of lower
values than those of the method proposed herein can be observed; the last estimate is even lower than the left non-dimensional depth under the bisector in Fig. (\ref{fig:SPFFS1}), which appears quite counter-intuitive. The threshold state that avoids choking is also plotted.
The corresponding non-dimensional specific energy $\Gamma_R^{RF}$ can
be easily computed, which allows the relative total head dissipation
on the step to be computed according to \cite{Bernetti2008, Rosatti2006, Rosatti2010, Cozzolino2011}: 

\begin{equation}
\frac{\Delta H^{RF}}{H_L}=\frac{H_L-H_R^{RF}}{H_L}=
\frac{\Gamma_L-\left[\Gamma_R^{RF}+\left(2/3\right)\,\delta\right]}{\Gamma_L}
\end{equation}

This is plotted in Fig. (\ref{fig:SPFFS2}) and shows a negative
relative total head dissipation on the order of $1 \cdot 10^{-1}$
for non-dimensional step heights of less than $0.5$. This result means that a
net gain of total head occurs in this case, which is clearly a
physical paradox. From a practical perspective, it is recommended to
avoid the RF method for the SPFFS case without choking to avoid an unphysical gain of energy crossing the step.

\subsection{Forward-facing step with choking (FFSCHK)}
\begin{figure}
\begin{center}
\includegraphics[width=1.0\textwidth]{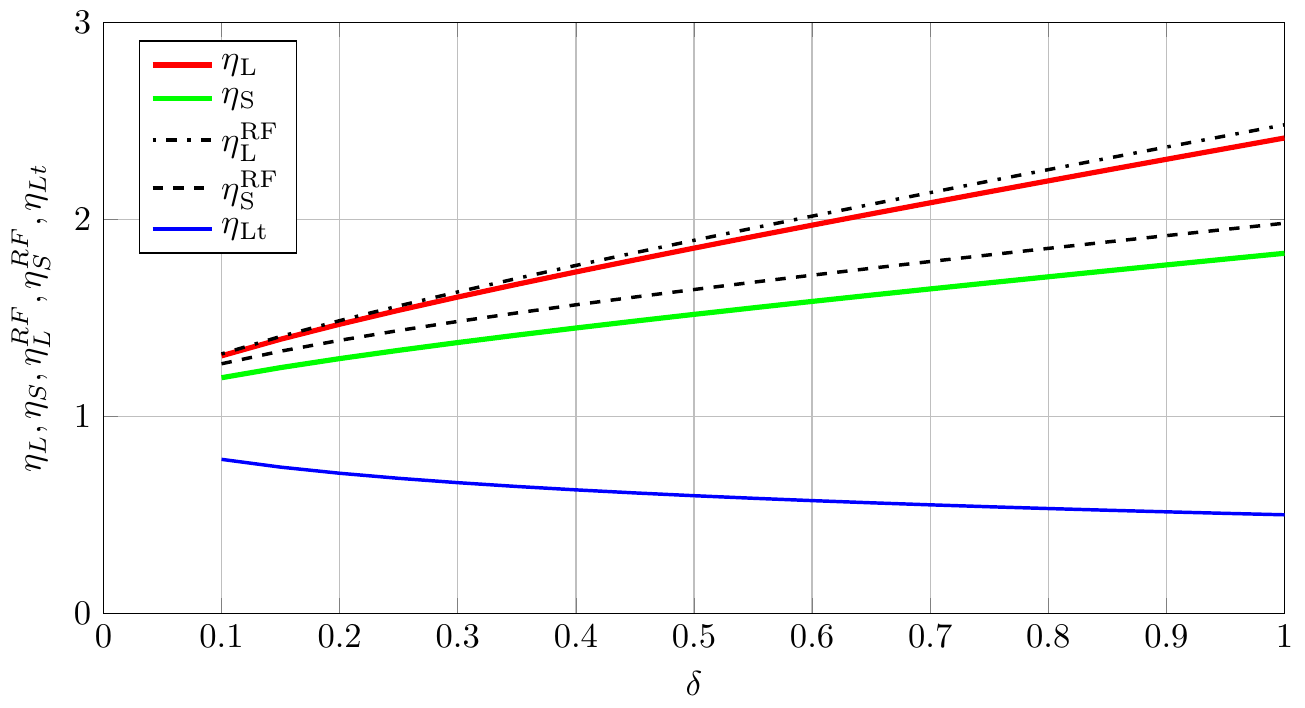}
\end{center}
\caption{FFSCHK: left and step non-dimensional depths {\vs} the
  non-dimensional step height in the case of choking}
\label{fig:FFSCHK1}
\end{figure}

Flow choking occurs when an attempt is made to prescribe the left
supercritical flow conditions and the left depth is not sufficiently
small (or similarly, the right Froude number is not sufficiently high
or the left specific energy of the flow is not sufficiently large), more specifically, when
\begin{equation} \label{eq:thresholdFFS}
\Gamma_L<1+\frac{2}{3}\,\delta
\end{equation}
In this case, the right state just downstream of the step shifts to
the critical state, and the left state just upstream of the step
shifts to the only subcritical state that has the minimum specific energy that is necessary and sufficient to pass the step:
\begin{equation} \label{eq:leftcrit}
\Gamma_R=1;\,\,\eta_R=1;\,\,\Phi_R=1
\end{equation}
\begin{equation}
\Gamma_L=1+\frac{2}{3}\,\delta\,\,\Rightarrow\eta_{L}
\end{equation}
The left subcritical depth is found using the inversion method, which is summarized in Tab. \ref{tab:eqtable}.
Once the couple of depths $\left(\eta_L,\,\eta_R=1\right)$ are known, the depth on the step, $\eta_S$, can be obtained from Eq. \eqref{eq:eta_S}, choosing the $-$ sign and $\Phi_R=1$. This value, which satisfies the condition $1=\eta_R \leq \eta_S \leq \eta_L$, is plotted in Fig. (\ref{fig:FFSCHK1}) as a function of the non-dimensional height of the step. In the same figure, the threshold value of the non-dimensional left depth $\eta_{Lt}$, as defined by Eq. (\ref{eq:thresholdFFS}), is plotted. If the prescribed left non-dimensional depth is greater than the threshold value, then flow choking occurs and the critical condition occurs at the  right side.

\begin{figure}
\begin{center}
\includegraphics[width=1.0\textwidth]{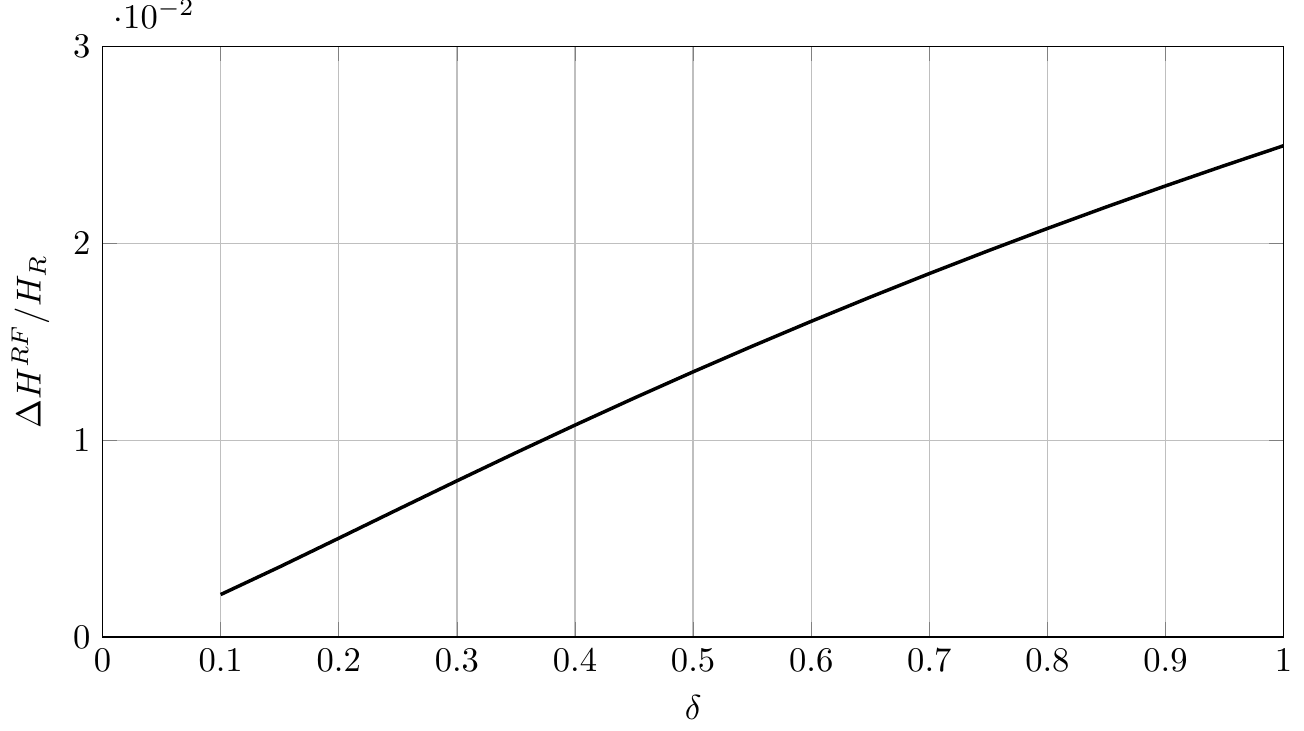}
\end{center}
\caption{FFSCHK: relative energy dissipation at the step according to RF \cite{Rosatti2006} and similar methods {\vs} the non-dimensional step height}
\label{fig:FFSCHK2}
\end{figure}

The left non-dimensional depth that satisfies the momentum balance
according to \cite{Bernetti2008, Rosatti2006, Rosatti2010,
  Cozzolino2011} is $\eta_L^{RF}$, and the corresponding
non-dimensional depth on the step is $\eta_S^{RF}=\eta_L^{RF} -
\delta/2$. Because the right state is the critical one, the value of  $\eta_L^{RF}$ is found from the non-dimensional momentum balance:

\begin{equation}
\frac{1}{3}\,\left(\eta_L^{RF}\right)^2+\frac{2}{3}\,\frac{1}{\eta_L^{RF}}-\frac{2}{3}\,\delta\,\left(\eta_L^{RF}-\frac{1}{2}\,\delta\right)=1
\end{equation}

The subcritical solution of this (3rd degree algebraic) equation is
$\eta_L^{RF}$. This value and the value of $\eta_S^{RF}$ are also
plotted in Fig. (\ref{fig:FFSCHK1}) as a function of the
non-dimensional height of the step. In this case, small differences
from the method proposed herein can be observed in the entire range of $\delta$.
The corresponding non-dimensional specific energy $\Gamma_L^{RF}$ can
be easily computed, which allows the relative total head dissipation
on the step to be computed according to \cite{Bernetti2008, Rosatti2006, Rosatti2010, Cozzolino2011}: 

\begin{equation}
\frac{\Delta H^{RF}}{H_R}=\frac{H_L^{RF}-H_R}{H_R}=
\frac{\Gamma_L^{RF}-\left[\left(2/3\right)\,\delta+1\right]}{\left[\left(2/3\right)\,\delta+1\right]}
\end{equation}

This is plotted in Fig. (\ref{fig:FFSCHK2}) and shows a relative total
head dissipation on the order of $1 \cdot 10^{-2}$ for non-dimensional
step heights of less than $1$. Such a dissipation, which is an increasing
function of the step height $\delta$, is a significantly different
result from the  conservative method proposed herein. As in the
previous cases, the order of magnitude of head losses in subcritical
flows is one order of magnitude lower than in supercritical
flows. Once choking occurs, these results are quite similar to the
case of SBFFS, as it is clearly managing two subcritical (one at the critical limit) depths.

\section{Conclusions}\label{sec:concl}
This work consist of a review of the SWE behaviour on a discontinuity
in bottom elevation. This review is focused on the differences between
the SWE scheme and real-world hydraulics on backward- and
forward-facing steps. In the classical limits of the SWE scheme, the pressure
distribution is hydrostatic on each vertical, and the flow velocity is
uniform on each vertical (by definition of SWE). Using the classical
results of \cite{LeFloch2011}, the conservation of generalized Riemann
invariants on the step is assumed. This means that over a stationary
step, the unit discharge and the total head of the flow are locally
constant. This reasoning allows for a complete and systematic review of
stationary flows over bottom steps strictly related to classical
hydraulics. It is explained that total head conservation is not in
contradiction with integral momentum conservation across the step,
with the
key element being the flow depth estimate {\it on\ }the step. Assuming
that this special depth exists, it is evaluated by imposing the total
force on the step as the result of a hydrostatic pressure distribution
on the step. Results are obtained for the entire range of stationary
cases, combining BFS or FFS with subcritical or supercritical flow
conditions and taking the occurrence of choking into account when
necessary. Exploring the same range of results using the RF
\cite{Bernetti2008, Rosatti2006} method, it is demonstrated that
subcritical flows on backward-facing steps and supercritical flows on
forward-facing step provide a couple of depths that produce a net gain
of energy across the step in stationary conditions. For this reason,
we highly recommend the use of total head conservation schemes such as that presented herein.
When head losses due to geometrical singularities must be taken into account, the introduction of specific, additional terms as suggested by classical handbooks (head losses expressed as $\xi \cdot U^2/\left(2 \, g\right)$, where $\xi$ is a proper non-dimensional coefficient that is expressed as a function of geometry) have to be considered.

\appendix

\section{Limits for the depth on the step}\label{sec:app_A}
In this appendix, the SBFFS case is chosen as an example. The procedure is completely similar for the entire range of cases
(both BFS and FFS) considered here. The system formed by Eqs.~\eqref{eq:enstepndext} and \eqref{eq:intmomstepndext} is considered.
Using simple algebra, it is possible to show the following relation:

\begin{equation}
\eta_S=\frac{\eta_L \, \eta_R \left(\eta_L^2 \, \eta_R  + \eta_L \, \eta_R^2 - 2 \right)}
{2\, \eta_L^2 \, \eta_R^2 - \left(\eta_L + \eta_R  \right)}
\end{equation}

For this case, where $\eta_L > \eta_R$, it is possible to demonstrate that:

\begin{equation}
\eta_S-\eta_L < 0; \quad \eta_S-\eta_R > 0
\end{equation}

that is, finally:

\begin{equation}
\eta_R < \eta_S <\eta_L
\end{equation}

Strictly similar proofs can be provided for the other cases in the investigated range.


\bibliographystyle{elsarticle-num} 



\end{document}